\documentclass[12pt]{article}
\usepackage{graphicx}
\usepackage{epsfig}
\usepackage{epstopdf}
\DeclareGraphicsExtensions{.pdf,.eps,.png,.jpg,.mps}

\def\lsim{\mathrel{\rlap{\lower3pt\hbox{\hskip0pt$\sim$}}
     \raise1pt\hbox{$<$}}}         
\def\gsim{\mathrel{\rlap{\lower4pt\hbox{\hskip1pt$\sim$}}
     \raise1pt\hbox{$>$}}}         

\usepackage{amsmath}
\usepackage{amsfonts}

\begin{document}
\begin{titlepage}

\centerline{\Large \bf A Spectral Model of Turnover Reduction}
\medskip

\centerline{Zura Kakushadze$^\S$$^\dag$\footnote{\, Zura Kakushadze, Ph.D., is the President of Quantigic$^\circledR$ Solutions LLC,
and a Full Professor at Free University of Tbilisi. Email: \tt zura@quantigic.com}}
\bigskip

\centerline{\em $^\S$ Quantigic$^\circledR$ Solutions LLC}
\centerline{\em 1127 High Ridge Road \#135, Stamford, CT 06905\,\,\footnote{\, DISCLAIMER: This address is used by the corresponding author for no
purpose other than to indicate his professional affiliation as is customary in
publications. In particular, the contents of this paper
are not intended as an investment, legal, tax or any other such advice,
and in no way represent views of Quantigic$^\circledR$ Solutions LLC,
the website \underline{www.quantigic.com} or any of their other affiliates.
}}
\centerline{\em $^\dag$ Free University of Tbilisi, Business School \& School of Physics}
\centerline{\em 240, David Agmashenebeli Alley, Tbilisi, 0159, Georgia}
\medskip
\centerline{(April 20, 2014; revised: July 24, 2015)}

\bigskip
\medskip

\begin{abstract}
{}We give a simple explicit formula for turnover reduction when a large number of alphas are traded on the same execution platform and trades are crossed internally. We model turnover reduction via alpha correlations. Then, for a large number of alphas, turnover reduction is related to the largest eigenvalue and the corresponding eigenvector of the alpha correlation matrix.
\end{abstract}

\bigskip
\medskip

\noindent{}{\bf Keywords:} hedge fund, alpha stream, crossing trades, transaction costs, portfolio turnover, correlation structure, large N limit
\end{titlepage}

\newpage

\section{Introduction and Summary}

{}With technological advances, hedge funds and similar investment vehicles can simultaneously trade multiple alpha streams.\footnote{\, Here ``alpha" means a real-life (as opposed to ``academic'') alpha, that is, any reasonable expected return on which one may wish to trade, that is, take risk. In fact, real-life alphas ({\em e.g.}, momentum strategies) often have sizable exposure to risk. Furthermore, there is no ``perfect'' risk model w.r.t. which one would hypothetically neutralize risk exposure of a portfolio. Otherwise, there would only be mean-reversion caused by temporary trading imbalances, which is evidently not the case in real life. Different time horizons provide different alpha (trading) opportunities.} One immediate question that arises is how to allocate capital to these alphas, or, mathematically speaking, how to determine weights with which the alphas should be combined. This is an optimization problem, whose solution depends on the precise optimization criterion as well other factors, such as if and how transaction costs are included and modeled.

{}The second issue is related to crossing trades between alphas. If, say, within a given hedge fund, Strategy A wants to buy \$1M of MSFT while Strategy B wants to sell \$1M of MSFT, it makes sense to cross this trade internally -- if the execution platform allows this, that is -- as opposed to going to the market as internal crossing amounts to substantial savings in transaction costs.\footnote{\, An illustrative discussion of internal crossing and its benefits can be found in (Kakushadze and Liew, 2014), including an explicit example of crossing trades between mean-reversion and momentum alphas. For a partial list of hedge fund literature, see, {\em e.g.}, Ackerman {\em et al} (1999), Agarwal and Naik (2000a, b), Amin and Kat (2003), Asness {\em et al} (2001), Brooks and Kat (2002), Brown {\em et al} (1999), Chan {\em et al} (2006), Edwards and Caglayan (2001), Edwards and Liew (1999a, b), Fung and Hsieh (1999, 2000, 2001), Kao (2002),
Liang (1999, 2000, 2001), Lo (2001), Racicot and Th\'{e}oret (2013), Schneeweis {\em et al} (1996).} When internal crossing is employed, portfolio turnover is reduced, so using even the simplest model for transaction costs in portfolio optimization requires accounting for turnover reduction.

{}As more and more alpha streams are combined, one expects that on average crossing should increase, and therefore the percentage of the dollar turnover with respect to the total dollar investment -- which percentage we refer to simply as ``turnover" -- is expected to decrease. In (Kakushadze and Liew, 2014) it was argued that turnover indeed decreases and converges to a non-vanishing limit. Generally, it is no easy feat to precisely describe internal crossing and turnover reduction. In a portfolio consisting of a large number of underlying tradable instruments ({\em e.g.}, stocks), precise details of internal crossing depend on the detailed portfolio position and trade data. The question then is if one can {\em model} expected turnover reduction -- on average, that is -- with some reasonable assumptions.

{}One observation is that the more correlated the trades are, the more correlated the alphas are, and the more correlated the trades are, the lower the internal crossing is expected to be. Therefore, while turnover reduction is not necessarily a simple ({\em e.g.}, linear) function of alpha correlations, it is clear that it is somehow related to them, so one can try to model turnover reduction based on alpha correlations, which are much more tractable than the position and trade data. The key observation in (Kakushadze and Liew, 2014) is that, when the number $N$ of alphas is large -- the ``Large $N$ Limit" -- the turnover reduction is indeed expected to simplify. In (Kakushadze and Liew, 2014) a simple model of turnover reduction was discussed, where one assumes a uniform pair-wise correlation $\rho$ between different alphas. Then, when the number of alphas is large, the portfolio turnover has a non-vanishing limit, which is linearly proportional to $\rho$.

{}In this note we propose a model of turnover reduction for a general alpha correlation matrix. We argue that in the large $N$ limit we can model turnover reduction using a spectral decomposition of the correlation matrix -- hence the ``Spectral Model" -- using its eigenvalues and eigenvectors. In this limit we have a non-trivial formula for turnover reduction, which is a generalization of (Kakushadze and Liew, 2014). The complementary factor-model based approach of (Kakushadze, 2014a) confirms our result here that turnover goes to a finite limit when $N$ is large.

{}To summarize, in this note we give an explicit spectral model of turnover reduction for a general alpha correlation matrix in the limit where the number of alphas is large. In this regime, this model can be used in estimating transaction costs, and in the problem of portfolio optimization with costs. The latter application of our model was implemented in (Kakushadze 2015a, 2015b).

{}The remainder of this paper is organized as follows. Definitions are in Section 2. Section 3 deals with positive-definiteness of the covariance (or correlation) matrix. Section 4 discusses our spectral model of turnover reduction and caveats. Our main result is given by Eqs. (\ref{T1}), (\ref{rhostar}), (\ref{T2}) and (\ref{star.prime}). We briefly conclude in Section 5.

\section{Definitions}

{}We have $N$ alphas $\alpha_i$, $i=1,\dots,N$. Each alpha is actually a time series $\alpha_i(t_s)$, $s=0,1,\dots,M$, where $t_0$ is the most recent time. Below $\alpha_i$ refers to $\alpha_i(t_0)$.

{}Let $C_{ij}$ be the sample covariance matrix of the $N$ time series $\alpha_i(t_s)$. Let $\Psi_{ij}$ be the corresponding correlation matrix, {\em i.e.},
\begin{equation}
 C_{ij} = \sigma_i~\sigma_j~\Psi_{ij}
\end{equation}
where $\Psi_{ii} = 1$.

{}To begin with, let us ignore trading costs. Alphas $\alpha_i$ are combined with some\footnote{\, For the following discussion it is not important what the actual values of these weights are or how they are computed ({\em e.g.}, via optimization, regression, {\em etc.}). We keep them arbitrary subject to the normalization condition $\sum_{i=1}^N |w_i|=1$. The weights $w_i$ can be negative (internal crossing).} weights $w_i$. The portfolio Profit and Loss (P\&L) is given by
\begin{equation}
 P = I~\sum_{i=1}^N \alpha_i~w_i
\end{equation}
where $I$ is the investment level (long plus short).

{}When linear trading costs are included, P\&L is given by
\begin{equation}
 P = I~\sum_{i=1}^N \alpha_i~w_i - L~D
\end{equation}
where $L$ includes all fixed trading costs (SEC fees, exchange fees, broker-dealer fees, {\em etc.}) and linear slippage. The linear cost assumes no impact, {\em i.e.}, trading
does not affect the stock prices. Also, $D = I~T$ is the dollar amount traded, and $T$ is the turnover. Let $\tau_i > 0$ be the turnovers corresponding to individual alphas $\alpha_i$. If we ignore turnover reduction resulting from combining alphas, then
\begin{equation}
 T = \sum_{i=1}^N T_i \equiv \sum_{i=1}^N \tau_i~|w_i|
\end{equation}
However, turnover reduction can be substantial and needs to be taken into account. To do this, we need to model turnover when $N$ alphas are combined. The basic idea behind such modeling is discussed in (Kakushadze and Liew, 2014), including the assumption (and its limitations) that internal crossing can be parameterized by correlations. Here, without repeating the arguments of (Kakushadze and Liew, 2014), we will discuss a model of turnover reduction based solely on the correlation matrix $\Psi_{ij}$. As in (Kakushadze and Liew, 2014), in this note our calculations are carried out in the framework where each alpha is traded in its own separate aggregation unit, and matching trades are crossed between separate aggregation units.

\section{``Fixing" Covariance Matrix}

{}Generally, the covariance matrix $C_{ij}$ can have the following undesirable properties. First, it can be (nearly) degenerate. Second, it may not be positive (semi-)definite (see footnote \ref{foot.neg.def}).

{}Let $V_i^{(a)}$ be $N$ right eigenvectors of $C_{ij}$ corresponding to its eigenvalues $\lambda^{(a)}$, $a=1,\dots,N$:
\begin{equation}
 C~V^{(a)} = \lambda^{(a)}~V^{(a)}
\end{equation}
with no summation over $a$. Let $U$ be the $N\times N$ matrix of eigenvectors $V^{(a)}$, {\em i.e.}, the $a$th column of $U$ is the vector $V^{(a)}$:
\begin{equation}
 U_{ij} \equiv V_i^{(j)}
\end{equation}
Let $\Lambda$ be the diagonal matrix of the eigenvalues $\lambda^{(a)}$:
\begin{equation}
 \Lambda_{ij} \equiv \delta_{ij} ~\lambda^{(j)}
\end{equation}
with no summation over $j$. Then
\begin{equation}
 C~U = U~\Lambda
\end{equation}
Note that, because $C$ is symmetric, $U$ can be chosen to be orthonormal: $U^T~U = 1$.

{}Let
\begin{equation}
 {\widetilde w} \equiv U^T~w
\end{equation}
Then the volatility $R$ is given by
\begin{equation}
 R = I~\sqrt{{\widetilde w}^T~\Lambda~{\widetilde w}} = I~\sqrt{\sum_{i=1}^N \lambda^{(i)}~{\widetilde w}_i^2}
\end{equation}
So, if $C_{ij}$ is not positive semi-definite, {\em i.e.}, if any of its eigenvalues $\lambda^{(a)}$ is negative, then the volatility $R$ is ill-defined. Also, if $C_{ij}$ is (nearly) degenerate, {\em i.e.}, if any of its eigenvalues $\lambda^{(a)}$ is zero (or small), then the corresponding linear combination of alphas given by
\begin{equation}\label{dir}
 \sum_{i=1}^N V_i^{(a)}~\alpha_i
\end{equation}
has zero (or small) contribution to the volatility $R$, thereby introducing an instability into the system.

{}Near degeneracy is caused by alphas that are almost 100\% correlated or anti-correlated and can be cured by simply removing such ``redundant" alphas:\footnote{\, The matrix $C_{ij}$ is degenerate if and only if the matrix $\Psi_{ij}$ is degenerate: $\det(C) = \det(\Psi)~\prod_{i=1}^N~\sigma_i^2$.} for each kept $\alpha_i$, each $\alpha_j$ ($j\not=i$) is removed so long as $\left|\Psi_{ij}\right| > \Psi_*$, where $0<\Psi_*<1$ is the upper bound on the modulus of the allowed correlations ({\em e.g.}, $\Psi_* = 0.9$). In the subsequent sections we will assume that $\left|\Psi_{ij}\right|\leq \Psi_* < 1$.

{}However, in practice, near degeneracy is usually caused by the fact that $M \ll N$. In fact, when $M < N$, only $M$ eigenvalues of $C_{ij}$ are non-zero, while the remainder have ``small" values, which can be positive or negative. These small values are zeros distorted by computational rounding.\footnote{\,  Actually, this assumes that there are no N/As in any of the alpha time series. If some or all alpha time series contain N/As in  non-uniform manner and the correlation matrix is computed by omitting such pair-wise N/As, then the resulting correlation matrix may have negative eigenvalues that are not ``small" in the sense used above, {\em i.e.}, they are not zeros distorted by computational rounding. The deformation method we discuss above can be applied in this case as well. Non-positive-definiteness of the original (undeformed) correlation matrix typically is not a dominant effect in the first principal component (see below) and turnover reduction; however, in practice one would typically use a positive-definite (deformed) correlation matrix and the deformation can have a sizable effect -- see Section 7 of (Kakushadze, 2014a) for illustrative empirical examples.\label{foot.neg.def}} In such cases, the solution is not to remove any alphas (as they are not necessarily ``redundant"), but to deform the covariance matrix so it is positive-definite.

\subsection{A Simple Method}

{}If one is interested in solving just the positive-definiteness problem, there are various ways of doing this. A simple method that does not require removing any alphas is as follows (Rebonato and J\"ackel, 1999). Suppose some eigenvalues $\lambda^{(a)}$ are negative or zero. Let
\begin{eqnarray}
 &&{\widetilde \Lambda} \equiv \mbox{diag}\left({\widetilde \lambda}^{(a)}\right)\\
 &&{\widetilde \lambda}^{(a)} \equiv \mbox{max}\left(\lambda^{(a)},\lambda_*\right),~~~a=1,\dots,N
\end{eqnarray}
where one chooses $\lambda_* > 0$.
Next, let
\begin{eqnarray}
 &&Z \equiv \mbox{diag}\left(z_i\right)\\
 &&z_i \equiv {C_{ii} \over {\sum_{j=1}^N U_{ij}^2~{\widetilde \lambda}^{(j)}}}
\end{eqnarray}
Finally, let
\begin{eqnarray}
 &&{\widetilde U} \equiv \sqrt{Z}~U\sqrt{{\widetilde \Lambda}}\\
 &&{\widetilde C} \equiv {\widetilde U}~{\widetilde U}^T
\end{eqnarray}
Note that ${\widetilde C}_{ij}$ is positive definite, and we have
\begin{equation}
 {\widetilde C}_{ii} = C_{ii}
\end{equation}
{\em I.e.}, this way we obtain a new positive-definite covariance matrix ${\widetilde C}_{ij}$ while preserving the diagonal elements of $C_{ij}$. Note that, instead of applying this method to the covariance matrix $C_{ij}$, one may choose to apply it directly to the correlation matrix $\Psi_{ij}$, as this method properly preserves the unit diagonal elements of $\Psi_{ij}$.

\section{Spectral Model}

{}The first observation is that, as we scale $T_i\rightarrow \zeta~T_i$, we must have $T\rightarrow \zeta~T$, where $\zeta > 0$. Next, let ${\widetilde V}^{(p)}_i$ be the eigenvectors of $\Psi_{ij}$ corresponding to the eigenvalues\footnote{\, Here we are assuming that, if need be, the method reviewed in Subsection 3.1 has been applied and all $\psi^{(p)}>0$. Furthermore, the basis of alphas $\alpha_i$ is taken ({\em i.e.}, the signs of $\alpha_i$ are chosen) such that $\sum_{i,j=1}^N\Psi_{ij} \equiv N^2\rho^\prime\equiv N(1 + (N-1){\overline \rho})$ is maximized (${\overline \rho}$ is the mean correlation).
Thus, consider the case with uniform correlations $\Psi_{ij}=\rho$, $i\neq j$, studied in (Kakushadze and Liew, 2014). In this case, in the large $N$ limit, the turnover reduction coefficient (see below) $\rho_* = {\overline \rho} = \rho$ (Kakushadze and Liew, 2014). However, if we flip the signs of {\em some} alphas $\alpha_i\rightarrow -\alpha_i$ (and then we must also flip the signs of the corresponding weights $w_i\rightarrow -w_i$), which does not change the portfolio turnover, the mean correlation ${\overline \rho}$ will no longer be equal $\rho$, hence the aforementioned choice of the basis for $\alpha_i$. We will discuss this point in more detail and give a precise prescription for fixing this basis below. For now we will just bear this in mind.\label{rho.prime}} $\psi^{(p)}$, $p=1,\dots,N$:
\begin{equation}
 \Psi ~{\widetilde V}^{(p)} = \psi^{(p)}~{\widetilde V}^{(p)}
\end{equation}
Let ${\widetilde U}_{ij}$ be the $N \times N$ matrix of eigenvectors ${\widetilde V}^{(p)}$, {\em i.e.}, the $p$th column of ${\widetilde U}$ is the vector ${\widetilde V}^{(p)}$:
\begin{equation}
 {\widetilde U}_{ij} \equiv {\widetilde V}_i^{(j)}
\end{equation}
${\widetilde U}$ can be chosen to be orthonormal: ${\widetilde U}^T~{\widetilde U} = 1$, which fixes the normalization of ${\widetilde V}^{(p)}$. Note that ${\widetilde V}^{(p)}$ form an orthonormal basis of $N$-vectors:
\begin{equation}
 \sum_{i=1}^N {\widetilde V}^{(p)}_i ~ {\widetilde V}^{(q)}_i = \delta_{pq}
\end{equation}
Let $\psi^{(1)} > \psi^{(2)}>\dots$ (so ${\widetilde V}^{(p)}$ are the principal components of $\Psi_{ij}$). Let
\begin{equation}
 {\widetilde T}^{(p)} \equiv\sum_{i=1}^N {\widetilde V}^{(p)}_i~T_i
\end{equation}
This is the basis in which $\Psi_{ij}$ is diagonalized:
\begin{equation}
 {\widetilde U}^T~\Psi ~{\widetilde U} = \mbox{diag}(\psi^{(p)})
\end{equation}
In this basis, the only relevant building blocks constructed solely from $T_i$ and $\Psi_{ij}$ are ${\widetilde T}^{(p)}$ and $\psi^{(p)}$, together with scalar invariants of $\Psi_{ij}$. Therefore, we have the following spectral model for the turnover:
\begin{equation}\label{T-sum}
 T = \kappa~\sum_{p=1}^N \psi^{(p)}~\left|{\widetilde T}^{(p)}\right| = \kappa~\sum_{p=1}^N \psi^{(p)}~\left|\sum_{i=1}^N {\widetilde V}^{(p)}_{i}~T_i\right|
\end{equation}
where $\kappa$ is a constant, which must be constructed from a scalar invariant of $\Psi_{ij}$. The only suitable scalar invariant is the trace.\footnote{\, Note that $\det(\Psi)$ is not suitable because $T$ is not expected to have a peculiar behavior when $\Psi_{ij}$ is nearly degenerate. Furthermore, only the trace-based scalar invariant reproduces the special case discussed below. Also, see below why relative coefficients in (\ref{T-sum}) do not change the end result.} Then $T$ is given by
\begin{equation}\label{model}
 T = {1\over\sqrt{{\mbox{Tr}(\Psi)}}}~\sum_{p=1}^N \psi^{(p)}~\left|\sum_{i=1}^N {\widetilde V}^{(p)}_{i}~T_i\right| =
 {1\over\sqrt{N}}~\sum_{p=1}^N \psi^{(p)}~\left|\sum_{i=1}^N {\widetilde V}^{(p)}_{i}~T_i\right|
\end{equation}
The power of $\mbox{Tr}(\Psi)$ and the overall coefficient are fixed as follows. Let all off-diagonal elements of $\Psi_{ij}$ be identical: $\Psi_{ij} = \rho$ ($i\not=j$). Also, let all $T_i$ be identical. Then, recalling that $\psi^{(1)}$ is the largest eigenvalue, we have
\begin{eqnarray}
 &&\sum_{i=1}^N {\widetilde V}^{(1)}_{i}~T_i = {1\over\sqrt{N}}~\sum_{i=1}^N T_i\\
 &&\sum_{i=1}^N {\widetilde V}^{(p)}_{i}~T_i = 0,~~~p>1\\
 &&\psi^{(1)}=1+(N-1)~\rho\\
 &&\psi^{(p)} = 1- \rho,~~~p>1\\
 &&T = {{1+(N-1)~\rho}\over N}~\sum_{i=1}^N T_i
\end{eqnarray}
which reproduces Eq. (19) in (Kakushadze and Liew, 2014).

{}The spectral model (\ref{model}) simplifies in the large $N$ limit. First, we fix the basis of alphas $\alpha_i$ as follows. Under the reflections $\alpha_i\rightarrow\eta_i\alpha_i$ (and, consequently, $w_i\rightarrow\eta_i w_i$), where $|\eta_i|=1$, we have $\Psi_{ij}\rightarrow \eta_i\eta_j\Psi_{ij}$, ${\widetilde V}_i^{(p)}\rightarrow \eta_i {\widetilde V}_i^{(p)}$, while $\psi^{(p)}$ are invariant. Therefore, we can always choose the basis such that all ${\widetilde V}^{(1)}_i\geq 0$. In what follows we always work in this basis. In the large $N$ limit, unless $T_i$ have a highly skewed distribution, the $p>1$ contributions in (\ref{model}) are suppressed\footnote{\, {\em E.g.}, in the uniform correlation case where $\Psi_{ij} = \rho$ ($i\not=j$), we have ${\widetilde V}^{(1)}_i = 1/\sqrt{N}$, while the rest of the eigenvectors have zero sums.} as ${\cal O}(1/N)$. Therefore, in the large $N$ limit the following simplified model is a good approximation:\footnote{\, In this regard, even if we allow nonuniform relative coefficients in the sum over $p$ in Eq. (\ref{T-sum}), in the large $N$ limit the subleading $p>1$ terms are suppressed and we still have (\ref{T1}).}
\begin{equation}\label{T1}
 T\approx {\psi^{(1)}\over\sqrt{N}}~\sum_{i=1}^N {\widetilde V}^{(1)}_{i}~T_i
\end{equation}
where $\psi^{(1)}$ is the largest eigenvalue of $\Psi_{ij}$, and ${\widetilde V}^{(1)}$ is the corresponding eigenvector (in the basis where all ${\widetilde V}^{(1)}_i\geq 0$) normalized such that
\begin{equation}
 \sum_{i=1}^N \left({\widetilde V}^{(1)}_i \right)^2 = 1
\end{equation}
Thus, assuming equal weights $w_i = 1/N$ with identical $\tau_i = \tau$, we have
\begin{equation}\label{Tequal}
 T \approx \rho_* ~\tau
\end{equation}
where
\begin{equation}\label{rhostar}
 \rho_* \equiv {\psi^{(1)}\over{N\sqrt{N}}}~\sum_{i=1}^N {\widetilde V}^{(1)}_{i}
\end{equation}
For a generic correlation matrix this quantity is constant with $N$ with high t-statistic. For an {\em illustrative}\footnote{\, We emphasize the adjective ``illustrative" for the reason that, because various hedge funds in this data do/did not all trade the same underlying instruments and also the corresponding time series are not 100\% overlapping (some hedge funds are dead, some are newer than others, {\em etc.}), it would not necessarily be correct to assume that their trades could be crossed. Therefore, we use this data only to {\em illustrate} various properties of the correlation matrix, and not to directly draw any conclusions about turnover reduction had these alpha streams actually crossed their trades.} example see Figure 1. The regression of $y$ over $x$ (without intercept) in Figure 1 has F-statistic over $1.5 \times 10^5$ (upper line with circles) and $5 \times 10^4$ (lower line with triangles). This confirms what was argued in (Kakushadze and Liew, 2014), that the turnover reduction based on the correlation matrix goes to a non-vanishing limit when $N$ is large, {\em i.e.}, $\rho_*$ does not vanish in this limit.

\subsection{Caveats}

{}The spectral model (\ref{model}) is exactly that -- a {\em model}. Its premise is that $T$ is built solely from building blocks constructed from $T_i$ and $\Psi_{ij}$. It is meant to work in the large $N$ limit and for generic configurations of $T_i$. For example, if all $T_i$ are zero except for $T_\ell$, $1\leq\ell\leq N$ ({\em i.e.}, $w_i = \delta_{i\ell}$, so $T_\ell = \tau_\ell$ and $T_i = 0$, $i\not=\ell$), then we expect $T=T_\ell$ as there is no internal crossing. Eq. (\ref{model}) does not have this property. In fact, one can attempt to construct such $T$ as follows. Let
\begin{equation}
 T = \sum_{p=1}^N B^{(p)}~\left|{\widetilde T}^{(p)}\right| = \sum_{p=1}^N B^{(p)}~\left|\sum_{i=1}^N {\widetilde V}^{(p)}_{i}~T_i\right|
\end{equation}
where $B^{(p)}$ are coefficients to be determined from the requirement that when $T_i = \delta_{i\ell}~\tau_\ell$ we have $T=T_{\ell}$:
\begin{equation}
 \sum_{p=1}^N B^{(p)} \left|{\widetilde V}^{(p)}_{i}~\right| = 1,~~~i=1,\dots,N
\end{equation}
This system of $N$ equations can be solved if the matrix $A_{ij}\equiv \left|{\widetilde V}^{(j)}_{i}~\right|$ is invertible. However, for a generic $\Psi_{ij}$ some of the coefficients $B^{(p)}$ will be negative. In any event, we will not pursue this direction here as our goal is turnover reduction in the large $N$ limit for generic configurations of $T_i$, which brings us to the next ``caveat".

{}For some $\Psi_{ij}$ some elements ${\widetilde V}^{(1)}_{i}$ in (\ref{T1}) can be small suppressing the contributions of the corresponding $\alpha_i$. We can remedy this via the following approximation:
\begin{equation}\label{T2}
 T\approx \rho_*~\sum_{i=1}^N T_i
\end{equation}
{\em I.e.}, in the sum over ${\widetilde V}^{(1)}_i~T_i$ in (\ref{T1}), $T_i$ are replaced by their cross-sectional average, and (\ref{Tequal}) is reproduced in the case where individual turnovers are uniform.

{}It might be tempting to replace $\rho_*$ by
\begin{equation}
 \rho^\prime \equiv {\psi_*\over N}
\end{equation}
where
\begin{equation}
 \psi_* \equiv {1\over N}\sum_{i,j=1}^N \Psi_{ij}
\end{equation}
is the least-squares solution to the approximate ``eigenvalue" equation:
\begin{eqnarray}
 && \Psi~{\overline V} \approx \psi_* ~{\overline V}\\
 &&\sum_{i=1}^N\left(\sum_{j=1}^N \Psi_{ij} - \psi_*\right)^2 \rightarrow {\mbox{min}}\label{minimize}
\end{eqnarray}
where the minimization in Eq. (\ref{minimize}) is w.r.t. $\psi_*$, and ${\overline V}_i \equiv 1/\sqrt{N}$ is the properly normalized unit vector. However, using $\rho^\prime$ can lead to underestimating turnover ({\em i.e.}, overestimating turnover reduction).\footnote{\, One may wish to use $\mbox{max}(\rho_*,\rho^\prime)$, their average or some other value between $\rho_*$ and $\rho^\prime$.} In the example of Figure 1, for the upper line (circles) we have $\rho_* \approx 0.282$ and $\rho^\prime\approx 0.252$, and for the lower line (triangles) we have $\rho_* \approx 0.127$ and $\rho^\prime\approx 0.110$.

{}In fact, there is a more precise relationship between $\rho_*$ and $\rho^\prime$. Thus, from
\begin{equation}
 \Psi_{ij} = \sum_{p=1}^N {\widetilde V}_i^{(p)}~{\widetilde V}_j^{(p)}~\psi^{(p)}
\end{equation}
we have
\begin{equation}\label{sum.v}
 \rho^\prime = {1\over N^2}~\sum_{i,j=1}^N \Psi_{ij} = {1\over N^2}~\sum_{p=1}^N \left[\sum_{i=1}^N{\widetilde V}_i^{(p)}\right]^2 \psi^{(p)} \approx {1\over N^2}~\left[\sum_{i=1}^N{\widetilde V}_i^{(1)}\right]^2 \psi^{(1)}
\end{equation}
where we have taken into account that in the large $N$ limit the $p>1$ terms in the sum are subleading. Combining (\ref{sum.v}) and (\ref{rhostar}), we get
\begin{equation}\label{star.prime}
 \rho_* \approx\sqrt{\rho^{(1)}~\rho^\prime}
\end{equation}
where $\rho^{(1)}\equiv \psi^{(1)}/N$. Eq. (\ref{sum.v}) makes it evident why in the large $N$ limit choosing the basis where all ${\widetilde V}_i^{(1)} \geq 0$ is equivalent to maximizing $\rho^\prime\approx {\overline \rho}$ (see footnote \ref{rho.prime}).

\subsection{Why Is All This Useful?}

{}In real-life trading, when one combines thousands of not-too-correlated alphas (with some weights) and trades the so-combined single ``unified" alpha on a single trading platform (as opposed to trading all these alphas on their own individual execution platforms), one gets an automatic bonus: internal crossing of trades between different alphas, hence turnover reduction, which can be substantial. Why is this important? Because the weights with which alphas are combined are determined via optimization (or a similar procedure) and including trading costs (and impact) into this optimization requires modeling turnover reduction, or else the effect of the costs would be (possibly grossly) overestimated, thereby resulting in a (possibly substantially) suboptimal alpha weights. Our main equations (\ref{T1}), (\ref{rhostar}), (\ref{T2}) and (\ref{star.prime}) model turnover reduction via the first principal component and the corresponding eigenvalue of the alpha correlation matrix, which is general. The model of turnover reduction in (Kakushadze and Liew, 2014) is a special simple case of our model where all pairwise correlations are uniform. This special case is an unrealistic toy model used in (Kakushadze and Liew, 2014) for the purpose of {\em illustrating} -- via its simplicity -- modeling turnover reduction via correlations and the existence of a nonvanishing limit for the turnover when $N$ goes to infinity (as opposed to a naive guess that turnover goes to zero in the large $N$ limit). However, the toy model of (Kakushadze and Liew, 2014) is just that -- a {\em toy} model. It is not designed for practical applications -- in real life correlations are not uniform. In contrast, our spectral model we give in this paper is designed precisely with practical applications in mind as it is applicable to a general (and realistic) alpha correlation structure. Put differently, if one uses Eq. (20) of (Kakushadze and Liew, 2014) in the general case, it is unclear what $\rho$ in that formula should be. What we have achieved here is that we give a simple explicit formula for this $\rho$ -- which we refer to as $\rho_*$ here -- via (\ref{rhostar}) in the general (that is, practically interesting) case. Furthermore, we cannot emphasize enough that our result here -- that $\rho_*$ is expressed via the first principal component and the corresponding eigenvalue -- only holds in the large $N$ limit; higher principal components are suppressed by powers of $1/N$, which are small when $N$ is sufficiently large. At finite $N$ there is no reason for such contributions to be small. Note that this is irrespective of whether we consider the general case or the toy model of (Kakushadze and Liew, 2014). In fact, Eq. (19) of (Kakushadze and Liew, 2014) expressly shows that unless $N$ is large, Eq. (20) of (Kakushadze and Liew, 2014), which is a special case of our model, does not hold.

{}One evident question arising in the context of the spectral model is out-of-sample stability. Generally, off-diagonal elements of a sample covariance (correlation) matrix are not expected to be too out-of-sample stable. Consequently, principal components of a correlation matrix inherit this instability. Nonetheless, the first principal component -- which happily is what our spectral model uses -- generally is the most out-of-sample stable, with higher principal components substantially more unstable. Prosaically, further mitigating the stability issue is the fact that turnover reduction is important if alphas have substantial turnover in the first instance, {\em i.e.}, the holding periods are short and, consequently, the relevant historical lookbacks are also short. With short lookbacks one recomputes quantities such as the correlation matrix and its first principal component on correspondingly high frequencies. In fact, there is yet another way of dealing with the stability issue if one is able to build a multi-factor risk model for alphas along the lines of (Kakushadze, 2014b), whereby instead of the sample correlation matrix one uses a constructed one, which by its very construction -- if such construction is possible in the first instance, that is -- is expected to be substantially more out-of-sample stable. All in all, in real life one works with what one has got and tries to do one's best with it, be it modeling turnover reduction, alpha covariance matrix, {\em etc.}

\section{Concluding Remarks}

{}The upshot is that -- just as in theoretical physics ('t Hooft, 1974) -- the large $N$ limit (Kakushadze and Liew, 2014) provides a powerful tool in quantitative finance. In this note we give an explicit spectral model of turnover reduction for a general alpha correlation matrix in the limit where the number of alphas is large. In this regime, this model can be used in estimating transaction costs, and in the problem of portfolio optimization with costs (Kakushadze, 2015a, 2015b). Our spectral model is expected to provide a good approximation for a generic distribution of individual alpha turnovers. In the large $N$ limit, the turnover reduction coefficient based on the spectral model does not appear to vanish but approaches a finite value. In (Kakushadze, 2014a) we further confirm the results of this paper by using a complementary factor model approach.

\newpage
\centerline{\epsfxsize 4.truein \epsfysize 4.truein\epsfbox{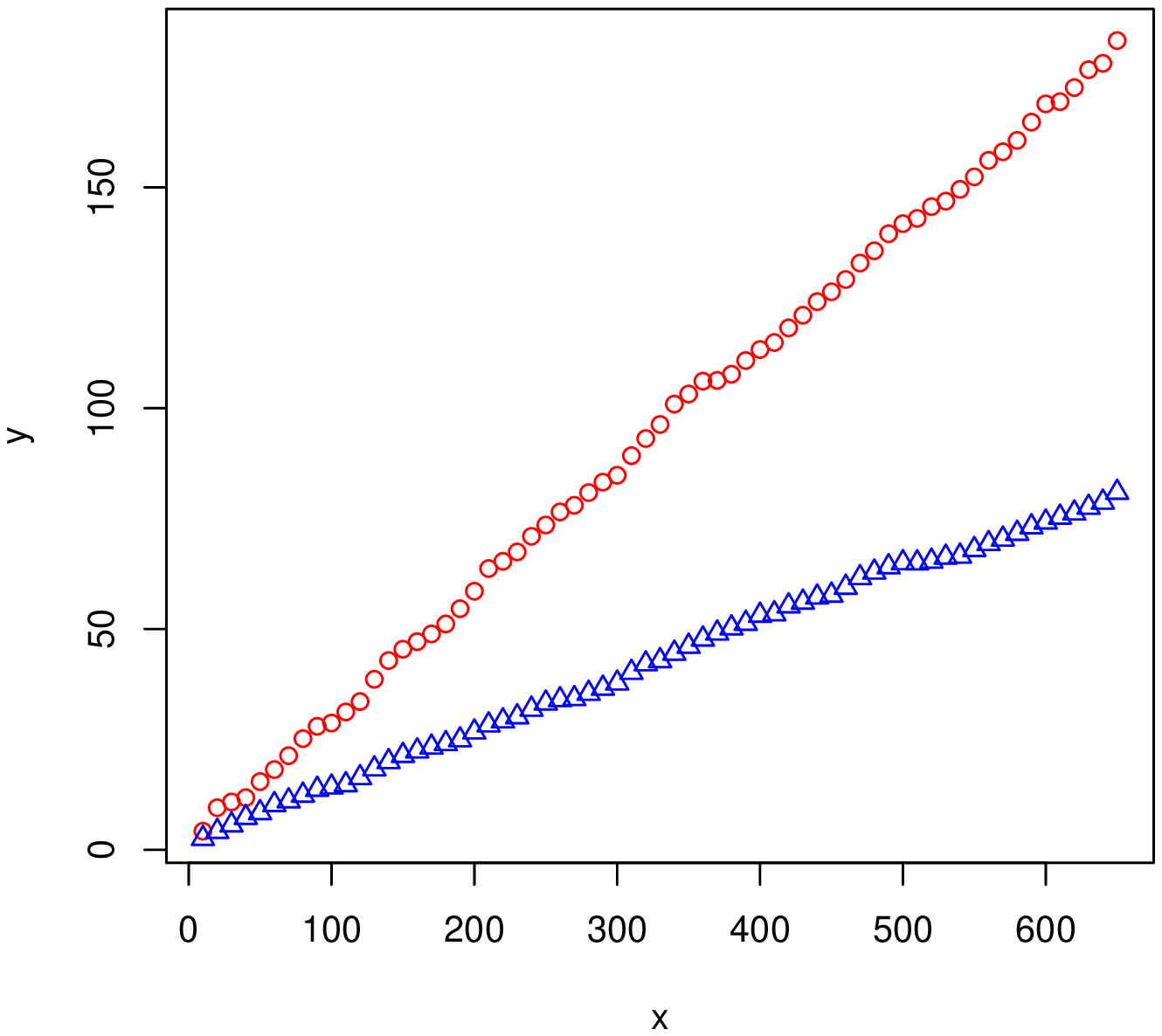}}
\noindent{\small {Figure 1. $x$-axis: $N$; $y$-axis: $\rho_* \times N$. This graph is based on the same Morningstar data for 1990-2014 for 657 hedge fund returns (HF) as Figure 1 in (Kakushadze and Liew, 2014). The upper line (circles) corresponds to the correlation matrix of the raw HF. The lower line (triangles) correspond to the correlation matrix of the residuals (plus the intercepts, which have no effect) of HF adjusted for RF (whose effect is small) and regressed over Mkt-RF and Fama-French risk factors SMB, HML, WML. Each correlation matrix is taken in the basis where its eigenvector corresponding to the largest eigenvalue has all nonnegative elements.}}


\begin{thebibliography}{99}

\bibitem{AMR} Ackerman, C., McEnally, R. and Revenscraft, D. (1999)
The Performance of Hedge Funds: Risk, Return and Incentives.
{\em Journal of Finance} 54(3): 833-874.

\bibitem{HF8} Agarwal, V. and Naik, N.Y. (2000a)
On Taking the ``Alternative" Route: The Risks, Rewards, and Performance Persistence of Hedge Funds.
{\em Journal of Alternative Investments} 2(4): 6-23.

\bibitem{HF9} Agarwal, V. and Naik, N.Y. (2000b)
Multi-Period Performance Persistence Analysis of Hedge Funds Source.
{\em Journal of Financial and Quantitative Analysis} 35(3): 327-342.

\bibitem{HF19} Amin, G. and Kat, H. (2003)
Stocks, Bonds and Hedge Funds: Not a Free Lunch!
{\em Journal of Portfolio Management} 29(4): 113-120.

\bibitem{HF12} Asness, C.S., Krail, R.J. and Liew, J.M. (2001)
Do Hedge Funds Hedge?
{\em Journal of Portfolio Management} 28(1): 6-19.

\bibitem{HF17} Brooks, C. and Kat, H.M. (2002)
The Statistical Properties of Hedge Fund Index Returns and Their Implications for Investors.
{\em Journal of Alternative Investments} 5(2): 26-44.

\bibitem{HF3} Brown, S.J., Goetzmann, W. and Ibbotson, R.G. (1999)
Offshore Hedge Funds: Survival and Performance, 1989-1995.
{\em Journal of Business} 72(1): 91-117.

\bibitem{HF20} Chan, N., Getmansky, M., Haas, S.M. and Lo, A.W. (2006)
Systemic Risk and Hedge Funds.
In: Carey, M. and Stulz, R.M. (eds.)
{\em The Risks of Financial Institutions}.
University of Chicago Press, Chapter 6, pp. 235-338.

\bibitem{HF13} Edwards, F.R. and Caglayan, M.O. (2001)
Hedge Fund and Commodity Fund Investments in Bull and Bear Markets.
{\em Journal of Portfolio Management} 27(4): 97-108.

\bibitem{HF4} Edwards, F.R. and Liew, J. (1999a)
Managed Commodity Funds.
{\em Journal of Futures Markets} 19(4): 377-411.

\bibitem{HF5} Edwards, F.R. and Liew, J. (1999b)
Hedge Funds versus Managed Futures as Asset Classes.
{\em Journal of Derivatives} 6(4): 45-64.

\bibitem{HF6} Fung, W. and Hsieh, D. (1999)
A Primer on Hedge Funds.
{\em Journal of Empirical Finance} 6(3): 309-331.

\bibitem{HF10} Fung, W. and Hsieh, D. (2000)
Performance Characteristics of Hedge Funds and Commodity Funds: Natural vs. Spurious Biases.
{\em Journal of Financial and Quantitative Analysis} 35(3): 291-307.

\bibitem{HF14} Fung, W. and Hsieh, D. (2001)
The Risk in Hedge Fund Strategies: Theory and Evidence from Trend Followers.
{\em Review of Financial Studies} 14(2): 313-341.

\bibitem{ZK} Kakushadze, Z. (2014a) Can Turnover Go to Zero?
{\em Journal of Derivatives \& Hedge Funds} 20(3): 157-176; http://ssrn.com/abstract=2444031.

\bibitem{AFM} Kakushadze, Z. (2014b) Factor Models for Alpha Streams.
{\em The Journal of Investment Strategies} 4(1): 83-109; http://ssrn.com/abstract=2449927.

\bibitem{AlphaOpt} Kakushadze, Z. (2015a)
Combining Alpha Streams with Costs.
{\em The Journal of Risk} 17(3): 57-78; http://ssrn.com/abstract=2438687.

\bibitem{AlphaWeights} Kakushadze, Z. (2015b)
Notes on Alpha Stream Optimization.
{\em The Journal of Investment Strategies} 4(3): 37-81; http://ssrn.com/abstract=2446328.

\bibitem{KL} Kakushadze, Z. and Liew, J.K.-S. (2014)
Is It Possible to OD on Alpha?
{\em The Journal of Alternative Investments} (forthcoming);\\
http://ssrn.com/abstract=2419415.

\bibitem{HF18} Kao, D.-L. (2002)
Battle for Alphas: Hedge Funds versus Long-Only Portfolios.
{\em Financial Analysts Journal} 58(2): 16-36.

\bibitem{HF7} Liang, B. (1999)
On the Performance of Hedge Funds.
{\em Financial Analysts Journal} 55(4): 72-85.

\bibitem{HF11} Liang, B. (2000)
Hedge Funds: The Living and the Dead.
{\em Journal of Financial and Quantitative Analysis} 35(3): 309-326.

\bibitem{HF15} Liang, B. (2001)
Hedge Fund Performance: 1990-1999.
{\em Financial Analysts Journal} 57(1): 11-18.

\bibitem{HF16} Lo, A.W. (2001)
Risk Management For Hedge Funds: Introduction and Overview.
{\em Financial Analysis Journal} 57(6): 16-33.

\bibitem{HFHF} Racicot, F.-\'{E}. and Th\'{e}oret, R. (2013)
The Procyclicality of Hedge Fund Alpha and Beta.
{\em Journal of Derivatives \& Hedge Funds} 19(2): 109-128.

\bibitem{RJ} Rebonato, R. and J\"ackel, P. (1999)
The most general methodology to create a valid correlation matrix for risk management and option pricing purposes. SSRN Working Paper, http://ssrn.com/abstract=1969689.

\bibitem{HF1} Schneeweis, T., Spurgin, R. and McCarthy, D. (1996)
Survivor Bias in Commodity Trading Advisor Performance.
{\em Journal of Futures Markets} 16(7): 757-772.

\bibitem{thooft} 't Hooft, G. (1974)
A Planar Diagram Theory For Strong Interactions.
{\em Nuclear Physics} B72(3): 461-473.

\end{thebibliography}
\end{document}